\documentclass[amsmath,amssymb,pra,10pt,twocolumn,superscriptaddress]{revtex4-1}
\usepackage{float}
\usepackage{amssymb,amsthm, amssymb, latexsym}
\usepackage{graphicx}
\usepackage{bm}       
\usepackage[english]{babel}
\usepackage{color}
\newcommand{\ket}[1]{\left|#1\right\rangle}
\newcommand{\bra}[1]{\left\langle#1\right|}

\newcommand{\be}{\begin{equation}}
\newcommand{\ee}{\end{equation}}
\newcommand{\bea}{\begin{eqnarray}}
\newcommand{\eea}{\end{eqnarray}}

\begin{document}
\title{Toward a measurement of the effective gauge field and the Born-Huang potential 
with atoms in chip traps}
\author{Zeynep Nilhan G\"urkan}
\affiliation{College of Engineering and Technology, American University of the Middle East, Egaila, 
54200, Kuwait}
\author{Erik Sj\"oqvist}
\affiliation{Department of Physics and Astronomy, Uppsala University, Box 516, 
Se-751 20 Uppsala, Sweden }
\author{Bj\"orn Hessmo}
\affiliation{Department of Applied Physics, Royal Institute of Technology, Se-106 91, Stockholm, Sweden}
\affiliation{Centre for Quantum Technologies, National
University of Singapore, 2 Science Drive 3, 117542 Singapore}
\author{Beno\^{i}t Gr\'{e}maud}
\affiliation{Aix Marseille Univ, Universit\'e de Toulon, CNRS, CPT, IPhU, AMUTech, Marseille, France}
\affiliation{Centre for Quantum Technologies, National
University of Singapore, 2 Science Drive 3, 117542 Singapore}
%
\today
\begin{abstract}{We study magnetic traps with very high trap frequencies where the spin is coupled to the motion 
of the atom. This allows us to investigate how the Born-Oppenheimer approximation fails and 
how effective magnetic and electric fields appear as the consequence of the non-adiabatic 
dynamics. The results are based on exact numerical diagonalization of the full Hamiltonian describing the coupling between the internal and external degrees of freedom. The position in energy and the decay rate of the trapping states correspond to the imaginary part of the resonances of this Hamiltonian are computed using the complex rotation method.}
\end{abstract}

\maketitle

\section{Introduction} 
Magnetic trapping is one of the workhorses for cold atom physics. A commonly used 
trap is the Ioffe-Pritchard trap \cite{pritchard1983}, in which the magnetic field is non-zero 
at the center to prevent Majorana transitions to scattering states. In most situations, 
it is highly desirable to obtain high trapping frequencies and gradients for fast thermalization 
and strong confinement of the atoms. One method to reach high frequencies is to miniaturize 
the magnetic trap by using microfabricated chips \cite{Fortagh2007}; a widely used setup for studying adiabatic dynamics in atomic systems \cite{Lesa}. This far most magnetic 
traps operate in a regime where the atomic spin can adiabatically follow local changes in 
the magnetic field. Before atoms are lost, the dynamics will be influenced by corrections to the adiabatic approximation. One such example would be where an atom orbits a current carrying wire. To describe losses and nonadiabatic effects in magnetic traps, it is convenient to work with waveguides to reduce the problem to two dimensions\cite{ AnglinSchmiedmayer, Franzosi, Hinds, SchmiedScrinzi}.

In this article, we analyze the situation in magnetic traps with very high trap frequencies, 
where the spin is coupled to the motion of the atom. This allows us to investigate how 
the Born-Oppenheimer approximation fails and how effective magnetic and electric fields 
appear as a consequence of the non-adiabatic dynamics 
\cite{Berry1990,mead1992,dalibard2011,goldman2014}. The results are based on exact numerical diagonalization of the full Hamiltonian describing the coupling between the internal and external degrees of freedom. Using the complex rotation method, the position in energy and the decay rate of the trapping states corresponding to the imaginary part of the resonances of this Hamiltonian are computed \cite{Buchl94}. 

\section{Magnetic trapping}

\subsection{Atoms in a magnetic field}
The Hamiltonian of an atom with mass $M$ and spin operator  ${\bf S}=(S_x,S_y,S_z)$ in a magnetic 
field ${\bf B}$, is given by  
\begin{eqnarray} 
H = \frac{{\bf p}^2}{2M} + 
\frac{g \mu_B}{\hbar} {\bf S} \cdot {\bf B} ,
\end{eqnarray}
where $\mu_B$ is the Bohr magneton and $g$ is the $g-$factor. We consider the $S=1$ 
case. By writing 
\begin{eqnarray}
{\bf B} = B (\sin{\alpha}\cos{\beta},\sin{\alpha}\sin{\beta},\cos{\alpha}),
\end{eqnarray}
where $B$, $\alpha$ and $\beta$ values are varying in space, the Hamiltonian reads 
\begin{eqnarray}
H & = & \frac{{\bf p}^2}{2M}  
\\
 & + & \frac{ \mu_B B_{tot}}{ \sqrt{2}}\left(
\begin{array}{ccc}
\sqrt{2}\cos\alpha & e^{-i\beta}\sin\alpha & 0 \\
e^{i\beta}\sin\alpha& 0 & e^{-i\beta}\sin\alpha \\
0 & e^{i\beta}\sin\alpha & -\sqrt{2} \cos \alpha
\end{array}
\right) 
\label{HS1},
\nonumber \end{eqnarray}
in the eigenbasis $\lvert -1\rangle, \vert 0\rangle, \vert 1\rangle $ of $S_z$.

When the Larmor frequency $\omega_L=\mu_B B_{tot}/\hbar$ is much larger than the 
typical frequency of the atomic motion, the fast spin dynamics decouple from the slow 
evolution of the center of mass of the atom. In the Born-Oppenheimer approximation, 
the state of the system $\Psi({\bf r})$ is written as 
\begin{equation} 
\vert\Psi ({\bf r}) \rangle \approx \psi_j ({\bf r}) \vert\chi_j ({\bf r})\rangle \label{ket} , 
\end{equation}
where  $\psi_j {({\bf r})}$ is the wave function and $\vert\chi_j ({\bf r}) \rangle$ is an eigenvector 
of the position dependent spin-part of the Hamiltonian $H$ in Eq. (\ref{HS1}). The effective 
potential seen by the atoms is just the associated eigenvalue $V_j ({\bf r})$, whose minimum, 
if existing, creates a trapping potential. In addition, since the local eigenstates 
$\vert \chi_j ({\bf r})\rangle$ are position dependent a vector potential and an additional 
scalar potential appear, for each component  
$\psi_j{({\bf r})}$:  
\begin{eqnarray} 
{\bf A}_j ({\bf r}) & = & - i \hbar\langle \chi_j ({\bf r})\vert \nabla \chi_j ({\bf r})\rangle , 
 \\
\Phi_j ({\bf r}) & = & \frac{\hbar^2}{2}\left[\langle\nabla\chi_{j}({\bf r})\vert
(1-\vert\chi_{j}({\bf r})\rangle \langle \chi_{j}({\bf r})\vert) \cdot \vert \nabla\chi_{j}({\bf r})\rangle \right].
\nonumber 
\end{eqnarray}
These quantities are well-known in molecular physics as Berry-Mead and Born-Huang 
potentials~\cite{mead1992}. In cold atomic gases, they appear in the situation of position 
dependent dark-states \cite{Rus}, allowing experimental realizations of artificial 
gauge fields \cite{Lin}. The eigenstates $\vert \chi_j ({\bf r})\rangle $ of the spin Hamiltonian 
are only defined up to a phase factor, which can be position dependent. From the preceding
expression, one readily obtains that the change $\vert \chi_j ({\bf r})\rangle\rightarrow 
e^{if({\bf r})/\hbar}\vert\chi_j ({\bf r})\rangle$  amounts 
to the gauge transformation:
\begin{eqnarray}
{\bf A}_j ({\bf r}) & \rightarrow & {\bf A}_j ({\bf r})+ \nabla f({\bf r}) , 
\nonumber \\ 
\Phi_j ({\bf r}) & \rightarrow & \Phi_j ({\bf r}) . 
\label{gauget}
\end{eqnarray}
In addition to these adiabatic terms, there are also couplings between the different eigenstates 
$\vert \chi_j ({\bf r})\rangle$, which appears as off-diagonal terms of the Hamiltonian in the 
$\vert \chi_j({\bf r})\rangle$ basis. They are precisely responsible for the Majorana losses 
 \cite{Suk,Bri}.

\subsection{Experimental set-up}
To observe the effects of the effective Berry-Mead and Born-Huang potentials, it is 
required to use magnetic traps with high trap frequencies. One efficient way to realize this 
is to use microfabricated magnetic traps, which are formed when the magnetic field from 
a small current-carrying wire is superimposed with an external bias field. The magnetic field 
from a thin and long wire along the $z$- axis is given by ${\bf B}_{\theta}(r) = 
\frac{\mu_0}{2 \pi} \frac{I_{\omega}}{r} {\bf e}_\theta$, where $\theta$ is the polar angle in 
cylindrical coordinates, $r$ is the distance from the wire, and $I_{\omega}$ is current. We assume a homogeneous 
external bias field ${\bf B}_{\textrm{bias}}= (\frac{B_0}{\sqrt{2}}, -\frac{B_0}{\sqrt{2}},B_z)$. 
Superimposing ${\bf B}_{\theta}(r)$ and ${\bf B}_{\textrm{bias}}$, we obtain, in cartesian coordinates, 
the total magnetic field
\begin{eqnarray}
{\bf B} & = & (\frac{B_0}{\sqrt{2}}-\frac{\mu_0}{2\pi}\frac{I_{\omega}y}{x^2+y^2}, 
-\frac{B_0}{\sqrt{2}}+\frac{\mu_0}{2\pi}\frac{I_{\omega}x}{x^2+y^2},B_z)
\nonumber \\
 & = & (\tilde{B}_x,\tilde{B}_y,\tilde{B_z}) .  
\end{eqnarray} 
The magnitude of ${\bf B}$ is given by 
\begin{equation} 
B^2 = B_z^2+B_0^2-\frac{\sqrt{2} B_0 \xi (x+y)}{x^2+y^2} + \frac{\xi^2}{x^2+y^2} 
\end{equation}
with $\xi=\frac{\mu_0 I_{\omega}}{2 \pi}$. This has a minimum at the position 
$x_0 = y_0 = \frac{\xi}{\sqrt{2} B_0}$, where weak-field seeking atoms can be trapped.

Introducing local coordinates $(\tilde{x},\tilde{y})$ around the minimum $(x_0,y_0)$, 
the $xy$-components of the magnetic field read:
\begin{eqnarray} 
\tilde{B}_x & = & \frac{\frac{B_0}{\sqrt{2}} 
((x_0+\tilde{x})^2+(y_0+\tilde{y})^2) - \xi (y_0+\tilde{y})}{(x_0+\tilde{x})^2+(y_0+\tilde{y})^2} , \\ \tilde{B}_y & = & \frac{-\frac{B_0}{\sqrt{2}} 
((x_0+\tilde{x})^2+(y_0+\tilde{y})^2)+\xi (x_0+\tilde{x})}{(x_0+\tilde{x})^2+(y_0+\tilde{y})^2} ,
\nonumber \end{eqnarray}
which to first order in $\tilde{x}$ and $\tilde{y}$ reduce to 
\begin{equation}
\tilde{B}_x \approx \frac{B_0^2}{\xi} \tilde{x}, \ \ \ \tilde{B}_y \approx -\frac{B_0^2}{\xi} \tilde{y} . 
\end{equation}
This approximation can be readily obtained from the linearization of ${\bf B}$ at the 
minimum of the potential: ${\bf B}\approx(G\tilde{x},-G\tilde{y},B_z)$, where $G$ is the gradient 
of the magnetic field in the $(\tilde{x},\tilde{y})$ plane, i.e., 
\begin{equation}
G = \frac{B_0^2}{\xi} = \frac{2 \pi}{\mu_0}\frac{B_0^2}{I_{\omega}} . 
\end{equation}


\subsection{Harmonic units}
As explained just above, weak-field seeking atoms can be trapped around the minimum of the 
magnetic field; more precisely, the effective trapping potential is proportional to:
\begin{eqnarray}
\mu_B B =\mu_B B_z \sqrt{1+\frac{G^2}{ B_z^2}\left(\tilde{x}^2+\tilde{y}^2\right)} , 
\end{eqnarray}
where the harmonic approximation leads an effective trap frequency (and harmonic length)
\begin{eqnarray} 
\omega_T = \sqrt{\frac{\mu_B B_z}{M}} \frac{G}{B_z} , 
\qquad \ell_T = \sqrt{\frac{\hbar}{M\omega_T}} . 
\end{eqnarray}
Using $\ell_T$ and $\omega_T$ as harmonic units for length and energy and by introducing 
the Larmor-type frequency $\omega_L= \mu_B B_z/ \hbar$, the Hamiltonian for a spin-$1$ 
reads 
\begin{eqnarray}
 & & \frac{H}{\hbar\omega_T}=\frac{{\bf p}^2}{2}  
\nonumber \\ 
 & & + \frac{V({\bf r})}{\sqrt{2}}\left( \begin{array}{ccc}
\sqrt{2}\cos\alpha & e^{-i\beta}\sin\alpha & 0 \\
e^{i\beta}\sin\alpha& 0 & e^{-i\beta}\sin\alpha \\
0 & e^{i\beta}\sin\alpha & -\sqrt{2} \cos \alpha 
\end{array}\right) , 
\end{eqnarray}
where
\begin{equation}
 V({\bf r})=\frac{\mu_B B_{tot}}{\hbar\omega_T}=\frac{\mu_B B_z}{\hbar\omega_T}
\sqrt{1+\frac{G^2\ell_T^2}{B_z^2}(x^2+y^2)} . 
\end{equation}
We are now using the notation $x$ and $y$ for the scaled position around the minimum, not 
to be confused with the original notation. In these rescaled units, one has ${\bf p}=-i\nabla$.

With the above definition, it follows that 
\begin{eqnarray}
\frac{\mu_B B_z}{\hbar\omega_T} & = & \frac{\omega_L}{\omega_T} \equiv \rho^{-2} , 
\nonumber \\
\frac{G^2\ell_T^2}{B_z^2} & = & \frac{\omega_T}{\omega_L} \equiv \rho^2 ,
\end{eqnarray}
in terms of which 
\begin{equation}
V({\bf r})=\frac{1}{\rho^2}\sqrt{1+\rho^2(x^2+y^2)} . 
\end{equation}
The ratio $\rho$ precisely compares the motional dynamics (frequency trap) to the spin dynamics 
(Larmor frequency). As one can see, the full dynamics of the problem depend only on this single 
dimensionless parameter. In the usual trapping situation, this is a small parameter, typically 
ranging from $10^{-3}$ to $10^{-1}$, telling that, (i) the Born-Oppenheimer approximation is 
valid, and (ii) the trap is almost harmonic $V({\bf r})\approx\frac{1}{\rho^2}+\frac{1}{2}(x^2+y^2)$. 
On the other hand, for $\rho\approx 1$, i.e.,  the timescale of the spin and the motional dynamics 
are comparable, deviations from the harmonic behavior are marked and all the effects beyond 
the Born-Oppenheimer approximation become important, in particular the Majorana's losses.

Finally, as functions of the magnetic field gradient $G$ (in $T/m$) and the bias field $B_z$ 
(in $\mu T $), one has for Rb$^{87}$: 
\begin{equation}
\begin{aligned}
\omega_T&=\sqrt{\frac{\mu_B}{M}}\frac{G}{\sqrt{B_z}}=8.135\times10^2\frac{G}{\sqrt{B_z}}\\
\nu_T&=\frac{\omega_T}{2\pi}=129.47\frac{G}{\sqrt{B_z}}\\
\frac{\omega_T}{\omega_L}&=\frac{\hbar}{\sqrt{\mu_BM}}\frac{G}{B_z^{3/2}} = 
9.25\times10^{-5}\frac{G}{B_z^{3/2}}.
\end{aligned}
\end{equation}
For a given value of $G$ and $B_z$, the dimensionfull 
values for frequencies and the decay rates 
are related to the numerical ones as follows:
\begin{equation}
\begin{aligned}
\nu_{\text{exp}}&=\nu_T\times E_{\text{num}} , \\
\frac{\Gamma_{\text{exp}}}{2\pi}&=\nu_T\times\Gamma_{\text{num}} , 
\end{aligned}
\end{equation}
see below for application.

\section{Beyond the Born-Oppenheimer approximation}
\subsection{Effective Hamiltonian}
As mentioned above, beside the vector potentials ${\bf A}_j$ and the Born-Huang scalar potentials 
$\Phi_j$, off-diagonal couplings between the components $\psi_j$ arise because of the motion 
of the atoms. From a mathematical point of view, this is nothing but saying that the operator 
$\frac{{\bf p}^2}{2}$ does not commute with the position dependent diagonalization of the 
spin Hamiltonian. More precisely, decomposing the state of the system $\vert \Psi({\bf r})\rangle$ 
in the eigenstates of the spin Hamiltonian, $\vert \Psi ({\bf r})\rangle  = \sum_j \psi_j ({\bf r}) 
\vert \chi_j({\bf r}) \rangle$, one can derive the effective Hamiltonian $H_{\textrm{eff}}$ acting 
on the wave functions $\psi_j ({\bf r})$. One has
\begin{equation}
\begin{aligned}
\vert \chi_{-}({\bf r})\rangle  & = \sin^2{\frac{\alpha}{2}}\vert 1 \rangle - 
\frac{1}{\sqrt{2}}\sin \alpha e^{i \beta}\vert 0 \rangle \\ 
 & + \cos^2{\frac{\alpha}{2}} e^{2 i \beta}\vert -1\rangle  , \\
\vert \chi_0({\bf r})\rangle & = -\frac{1}{\sqrt{2}} \sin \alpha \vert 1\rangle+\cos \alpha 
e^{i \beta}\vert 0\rangle \\ 
 & + \frac{1}{\sqrt{2}} \sin \alpha e^{2 i \beta}\vert -1 \rangle , \\
\vert \chi_+({\bf r})\rangle & =  \cos^2{\frac{\alpha}{2}} \vert 1 \rangle + 
\frac{1}{\sqrt{2}} \sin \alpha e^{i \beta}\vert 0 \rangle \\ 
 & + \sin^2{\frac{\alpha}{2}} e^{2 i \beta}\vert -1\rangle,
\end{aligned}
\end{equation}
where the $\vert m \rangle$ are the Zeeman states. The associated eigenvalues are $-V({\bf r})$, 
$0$ and $+V({\bf r})$, respectively.

Assuming that the magnetic field simply reads $(Gx,-Gy,B_z)$, one obtains 
\begin{equation}
\begin{aligned}
\tan{\beta}&=-y/x=-\tan{\theta}\\
\tan{\alpha}&=\frac{G}{B_z}\sqrt{\tilde{x}^2+\tilde{y}^2}=\frac{G\ell_T}{B_z}\sqrt{x^2+y^2}=\rho\sqrt{x^2+y^2} . 
\end{aligned}
\end{equation}
In the adiabatic limit, 
i.e., $\rho\rightarrow 0$, $\alpha\rightarrow 0$, the trapping state is
$\vert \chi_+ ({\bf r})\rangle \approx \vert S_z=1\rangle$, 

In harmonic units, $H_{\mathrm{eff}}$ acting on the vector $(\psi_+, \psi_0,\psi_-)$ 
reads formally as follows:
\begin{equation}
 H_{\mathrm{eff}}=\left(
    \begin{array}{ccc}
     h_{++} & h_{+0} & h_{+-} \\
       h_{0+} & h_{00} & h_{0-} \\
 h_{-+} & h_{-0} & h_{--} 
    \end{array}
  \right)+
  \left(\begin{array}{ccc}
     V({\bf r}) & 0 & 0 \\
       0 & 0 & 0 \\
 0 & 0 & -V({\bf r})
    \end{array}
  \right)
\end{equation}
The diagonal entries take the form 
\begin{eqnarray} 
h_{jj} = \frac{1}{2} ({\bf p}-{\bf A}_j)^2 + \Phi_j , 
\end{eqnarray}
where 
\begin{equation}
 \begin{aligned}
  {\bf A}_{+}({\bf r})&= \left( 1- \frac{1}{\gamma} \right)\frac{{\bf e}_{\theta}}{r} , \\
  {\bf A}_0({\bf r})&= \frac{{\bf e}_{\theta}}{r} , \\
  {\bf A}_{-}({\bf r})&= \left( 1+\frac{1}{\gamma} \right)\frac{{\bf e}_{\theta}}{r} , \\
  \Phi_{-}({\bf x})&=2\Phi_{0}({\bf x})=\Phi_+({\bf x})=\frac{\rho^2(1+\gamma^2)}{4 \gamma^4}
 \end{aligned}
\end{equation}
with $r=\sqrt{x^2+y^2}$ and $\gamma = \gamma (r) = \sqrt{1+\rho^2r^2}$. 

The off-diagonal entries of $H_{\mathrm{eff}}$ are 
\begin{equation}
\begin{aligned}
h_{+0} &= -\frac{\sqrt{2} \rho}{2r \gamma^2}
\left(- r \frac{\partial}{\partial r}-i \gamma \frac{\partial}{\partial \theta}\right) 
+\frac{\sqrt{2} \rho}{2r \gamma^4}(\gamma^3-\gamma^2+1)  ,\\
h_{+-} & = \frac{r^2 \rho^4}{4 \gamma^4} , \\
h_{0+} & = -\frac{\sqrt{2} \rho}{2 r \gamma^2}
\left( r \frac{\partial}{\partial r}-i \gamma \frac{\partial}{\partial \theta}\right)
+\frac{\sqrt{2} \rho}{2r\gamma^4}(\gamma^3-1) , \\ 
h_{0-} &=\frac{\sqrt{2} \rho}{2 r \gamma^2}
\left( r \frac{\partial}{\partial r}+i \gamma \frac{\partial}{\partial \theta}\right)
+\frac{\sqrt{2} \rho}{2r \gamma^4}(\gamma^3+1) , \\
h_{-+} & = \frac{r^2 \rho^4}{4 \gamma^4} , \\
h_{-0} &=- \frac{\sqrt{2} \rho}{2r \gamma^2}
\left( r \frac{\partial}{\partial r}-i \gamma \frac{\partial}{\partial \theta}\right)
+\frac{\sqrt{2} \rho}{2r \gamma^4}(\gamma^3+\gamma^2-1) .\\
\end{aligned}
\end{equation}
$H_{\mathrm{eff}}$ is Hermitian with respect to the scalar product $\langle f(r,\theta)\vert g(r,\theta)\rangle = 
\iint rdr\,d\theta\, f^*(r,\theta)g(r,\theta)$. 

\subsection{Numerical implementation}
The Hamiltonian $H_{\mathrm{eff}}$ is invariant under spatial rotations, therefore the eigenstates 
can be written as $e^{im\theta}(\psi_+(r), \psi_0(r),\psi_-(r))$. For each value of $m$, the resulting 
Hamiltonian $H_m$ only depends on the radial coordinate $r$. Since the value of the functions 
$\psi_j(r)$ at $r=0$ is not fixed by any boundary condition, one uses a discretization scheme that 
does not contain the point $r=0$, namely the grid points are $r_n=(n+1/2)\Delta r$, for $n$ 
ranging from $0$ to a maximum value $N$. $1/\Delta r$ fixes the number of grid points per 
harmonic length, whereas $N\Delta r$ corresponds to the size of the system in harmonic length 
units. The Hamiltonian $H_m$ is Hermitian for the scalar product $\langle F\vert G\rangle = 
\int r dr  \left(f_+^*(r)g_+(r)+f_0^*(r)g_0(r)+f_-^*(r)g_-(r)\right)$, where $F=(f_+(r), f_0(r),f_-(r))$, 
such that it is the discretization of the equation $rH_m{\bf \psi}=Er{\bf \psi}$ that leads to
a generalized eigenvalue problem $AX=EBX$, where $A$ and $B$ are $(3N+3)\times(3N+3)$ 
Hermitian matrices, $B$ being positive definite.

Neglecting the off-diagonal coupling, the effective potentials seen by $\psi_0$ and $\psi_-$ 
components being $0$ and $-V({\bf r})$, the corresponding eigenstates are not bound states 
but scattering states, whereas they are bound states for the $\psi_+$ component. The off-diagonal 
coupling allows this bound states to decay to the scattering channels, which, from a mathematical 
point of view, becomes resonances, i.e., complex poles of the Green function $G(z)=1/(z-H)$.   
The complex rotation method is appropriate to compute directly the properties of these 
resonances (energy, width). Its properties rely on mathematical properties of the analytic
continuation of the Green's function in the complex plane \cite{Balslev71,Ho83}. A review of 
its application to atomic physics can be found in \cite{Buchl94}. 

The method is implemented in our case by making the radial coordinate complex:
$r\rightarrow e^{i\phi}r$ and $\frac{\partial}{\partial r} \rightarrow 
e^{-i\phi}\frac{\partial}{\partial r}$, where $\phi$ is a real parameter (the rotation angle). 
The matrix representations of the Hamiltonian then become complex symmetric, 
but are no more Hermitian. The fundamental properties of the complex spectra are :

\begin{itemize}
\item The bound states, if any, are still on the real axis.
\item The continua are rotated by an angle $2\phi$ on the lower-half complex
plane, around their branching point. 
\item Each complex eigenvalue $E_j$ gives the properties of one resonance, i.e., 
the energy is the real part of $E_j$, and the width is two times the negative
of the imaginary part. The complex eigenvalues are independent of $\phi$,
provided that they are not covered by the continua.
\end{itemize}

Note that in the present case, the branching point associated to $\psi_0$ is located at 
$E=0$, whereas the one associated to $\psi_-$ is formally at $E=-\infty$, since 
$-V(r)\approx -r/\rho$ for large $r$. The scattering states for a linear potential oscillate faster 
for large distance, such that they cannot be accurately described within a discretization 
scheme. On the other hand, the overlap with the bound states at very large distance is 
negligible, so that the exact behavior barely impacts the position and the width of the 
resonance. Therefore to avoid numerical artifacts and increase the numerical accuracy, 
we replace the anti-trapping  potential for $\psi_-$ by 
\begin{equation}
V_{\mathrm{scatt}}(r)=\frac{1}{\rho^2}\sqrt{1+\rho^2\frac{r^2}{1+h^2r^2}},
\end{equation}
where $h$ is a small parameter, typically $h\approx10^{-2}$. For $hr\ll1$, then 
$V_{\mathrm{scatt}}(r)\approx V(r)$, whereas for $hr\gg 1$, then $V_{\mathrm{scatt}} (9r) 
\approx 1/h\rho$, such that the scattering states have a well defined wavelength for large 
$r$ values. We have numerically checked that the results presented here are insensitive to 
the actual $h$ value. 

\subsection{Results}
\begin{figure}[H]
\centerline{\includegraphics[width=7cm]{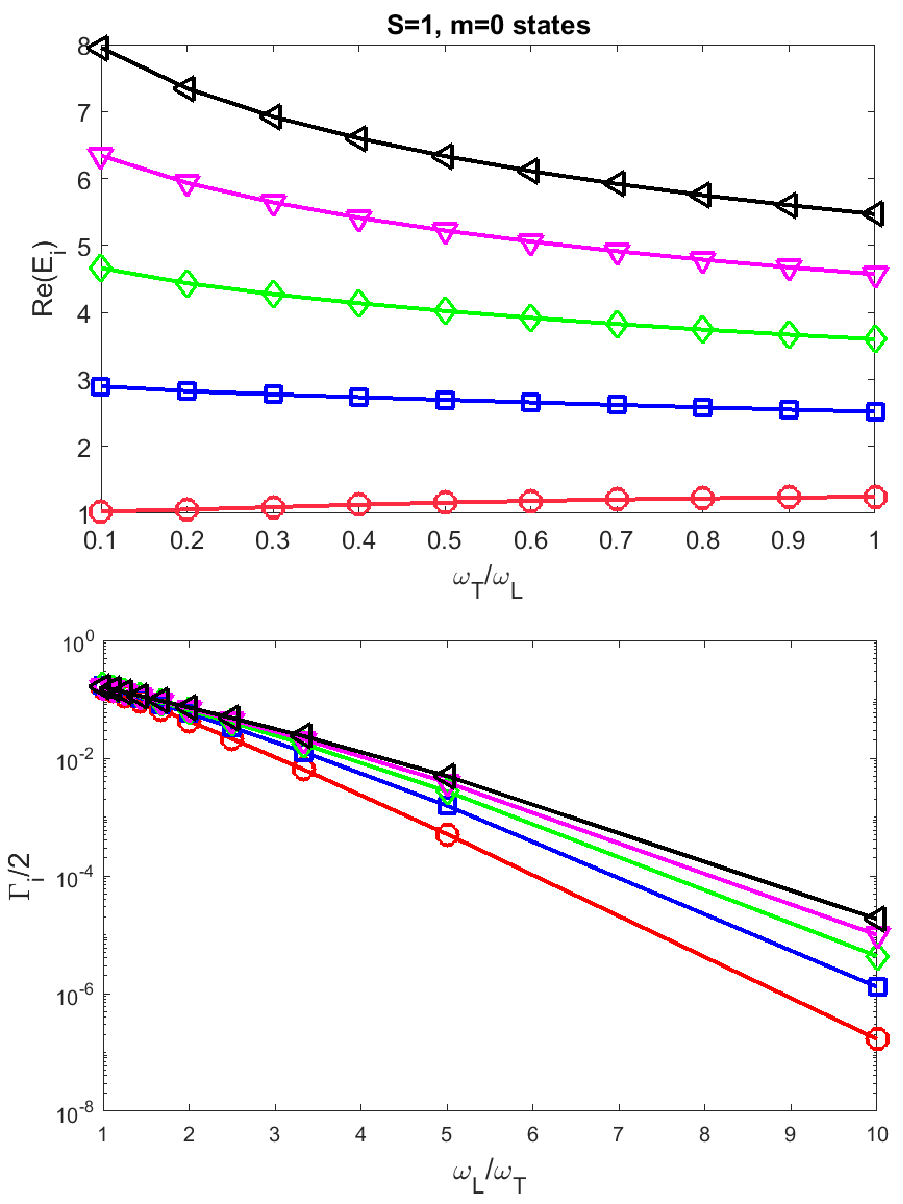}
}
\caption{(color online) Properties of the $m=0$ resonances of the Hamiltonian 
$H_{\mathrm{eff}}$: positions in energy as functions of $\omega_T/\omega_L$ (top plot) and decay rates as functions of $\omega_L/\omega_T$ (bottom plot). A given symbol and color correspond to the same state for the
two plots.}
\label{m0}
\end{figure}

\begin{figure}[H]
\centerline{\includegraphics[width=7cm]{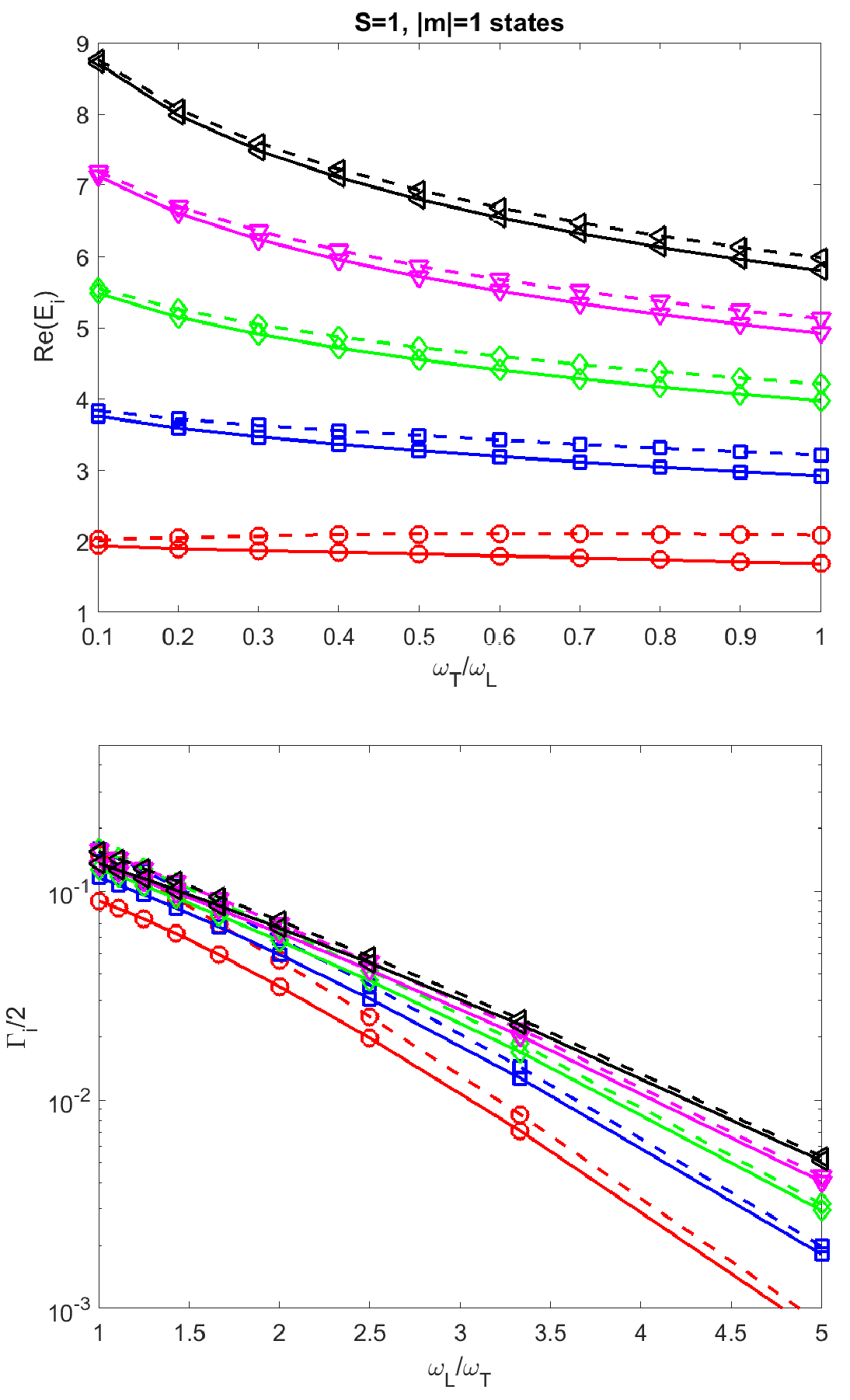}}
\caption{(color online) Properties of the $m=\pm1$ resonances of the Hamiltonian 
$H_{\mathrm{eff}}$: positions in energy as functions of $\omega_T/\omega_L$ (top plot)
and decay rates as functions of $\omega_L/\omega_T$ (bottom plot). The dashed line is for $m=-1$ and the solid line is for $m=1$. A given symbol and color correspond to the same state for the
two plots.}
\label{m1}
\end{figure}

The properties of the resonances (position in energy and width) are displayed, as functions of 
$\omega_T/\omega_L$, in Fig.~\ref{m0} for the $m=0$ states and in Fig.~\ref{m1} 
for the $m=\pm1$ states. In both cases, the zero of energy corresponds to the bottom of the 
trapping state, i.e., corresponding to a global shift of $-\frac{1}{\rho^2}$ of the eigenenergies 
of $H_{\mathrm{eff}}$, such that in the adiabatic limit $\rho\rightarrow0$, the energies directly 
correspond to the harmonic oscillator levels. This is clearly seen on the left part of the top plots. 
On the contrary, for $\rho\approx 1$ (right part), the energy levels differ from the harmonic 
one, in particular, the difference in energy $E_{n+1}-E_{n}$ gets smaller with higher $n$ reflecting 
the linear behavior of the trapping potential $V(r)$ at large $r$. The behavior in the adiabatic 
regime can be obtained from the perturbation expansion of $h_{++}$ with respect to ${\bf A}_{+}$ 
and $\phi_+$. More precisely, for a fixed value of $m$, $h_{++}$ reads:
\begin{multline}
    h_{++}=h_{++}^{(0)}-\frac{m}{r^2}\left(1-\frac{1}{\gamma}\right)+\\
\frac{1}{4r^2}\left(\frac{3\gamma^4-4\gamma^3-2\gamma^2+1}{\gamma^4}\right) 
\end{multline}

where  $h_{++}^{(0)}=\frac{p^2}{2}$ when $ \rho \rightarrow 0$.
The  second term arises from the gauge field ${\bf A}_{+}$ and results in a energy 
splitting among the $\pm m$ states, which is clearly observed in Fig.~\ref{m1}. A Taylor expansion 
of these two terms at small distances $\rho r\ll 1$ leads respectively to:
\begin{equation}
  \begin{aligned}
    -\frac{m}{r^2}\left(1-\frac{1}{\gamma}\right)&\approx-m\left(\frac{\rho^2}{2}-\frac{3}{8}\rho^4r^2\right)\\
    \frac{1}{4r^2}\left(\frac{3\gamma^4-4\gamma^3-2\gamma^2+1}{\gamma^4}\right)&\approx\frac{\rho^2}{2}-\frac{5}{8}\rho^4r^2
  \end{aligned}
\end{equation}
The two terms proportional to $\rho^2$ correspond to an energy shift whereas the two terms 
proportional to $\rho^4r^2$ correspond to modification of the trap frequency. However, the 
preceding formula cannot be compared directly to the numerical results since at large distances 
$\rho r\gg 1$ , the two last terms in $h_{++}$ leads to an effective centrifugal potential 
$(-m+\frac{3}{4})r^{-2}$, which is independent of $\rho$. This shows that, although ${\bf A}_{+}$ 
and $\phi_+$ are 
small perturbations to the $h_{00}$, the resulting energy shift has to be computed using their 
exact expressions, not just their Taylor expansion around $r=0$. Furthermore, in the adiabatic regime, the decay rates can be obtained from the Fermi golden 
rule. Assuming that one can approximate the scattering states as plane waves $\psi_0(r) 
\approx e^{ikr}$, i.e., neglecting the fact that $r$ is a radial coordinate, the decay rates are 
proportional to the modulus square of the Fourier transform $\vert \psi^{(n)}_{+} (k) \vert ^2$ of the 
harmonic trap wave functions, taken at the $k$ corresponding to the energy of the bound state, 
i.e., such that $\frac{k^2}{2}=E_n+\frac{1}{\rho^2}$. At first order the decay is dominated by 
the Gaussian decay of the wave function $e^{-r^2/2}$, such that one has:
\begin{equation}
 \Gamma\approx e^{-k^2}=e^{-2E_n-2/\rho^2}.
\end{equation}
Therefore, one predicts a linear behavior of $\ln \Gamma$ as functions of $\rho^{-2} = 
\omega_L/\omega_T$. This can be seen for both $m=0$ and $m=\pm 1$ states 
 (Figs. \ref{m0} and \ref{m1}, bottom plots).

\begin{figure}[H]
\centerline{\includegraphics[width=7cm]{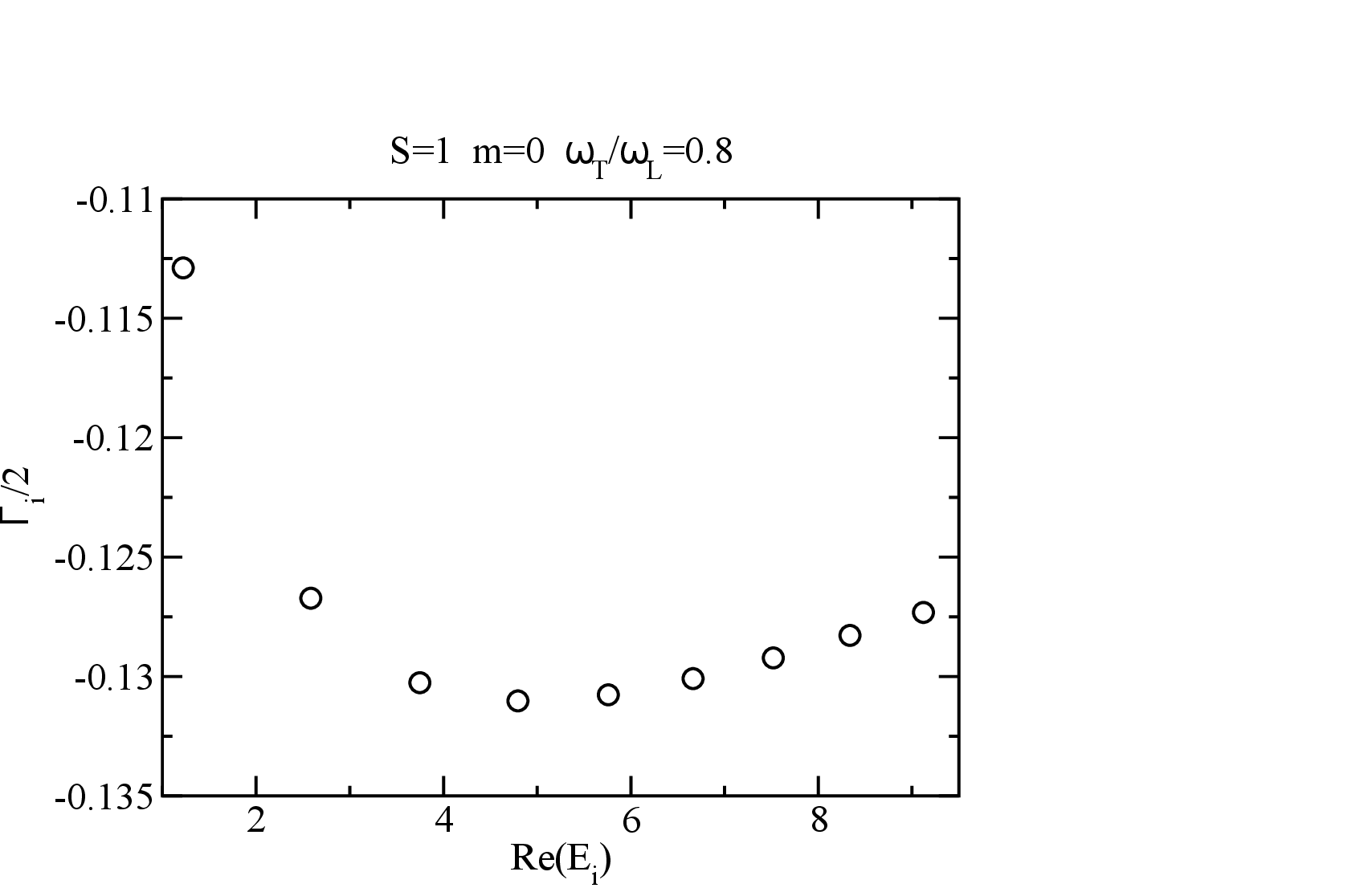}}
\caption{Resonances in the complex energy plane for a fixed value 
of $\rho^2=\frac{\omega_T}{\omega_L}=0.9$. As one can see, around an energy value that depends on $\rho$, $ ( E \approx 5 $ for $\rho^2=0.9 )$, the imaginary part of the resonances attains a (negative) minimum value, corresponding therefore to a maximum value of the decay rate.}
\label{resonances}
\end{figure}

The behavior, at fixed $\rho$, is slightly more complicated since one has to 
take into account the exact shape of the harmonic functions, in particular to explain that for 
small $\rho$, the decay rates increase with the principal quantum number. We have checked on a simpler model that it is indeed a generic behavior: for a fixed value 
of $\rho$, the decay rates depict a maximum around an energy (which depends on $\rho$), 
see Fig.~\ref{resonances}.

\begin{figure}[H]
\centerline{\includegraphics[width=7 cm]{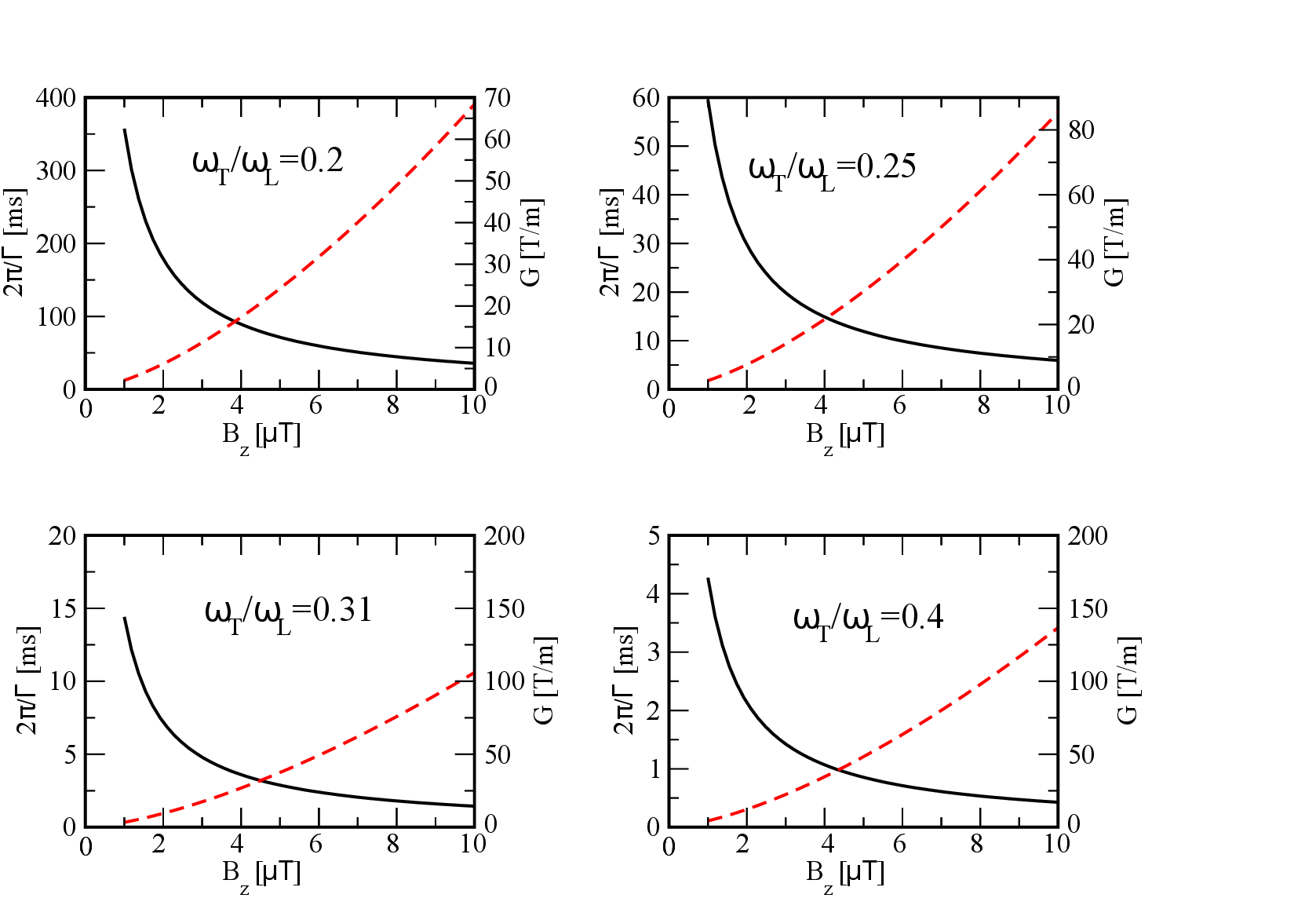}}
\caption{(color online) Lifetime $2\pi/\Gamma$  (solid black line, left axis) and the gradient 
$G$ (red dashed line, right axis) as functions of the bias field along the $z$ axis, for the four 
different values of $\omega_T/\omega_L$ depicted in the Table~\ref{exptab}. For instance, 
for a value of $\omega_T/\omega_L=0.2$ (top-left plot), for a bias field value of $5 \mu T$ , 
the life-time is $\approx70$ms, whereas $G$ is $\approx25 T/m $. For this value, the trap 
frequency is $\approx14.5$kHz. The energy splitting between the $\pm 1$ states is 
$2.2$kHz.}
\label{resonance}
\end{figure}

From the numerical point of view, Table~\ref{exptab} summarizes the expected decay rate 
of the $m=0$ ground state and the energy splitting between the first two $m=\pm1$ states 
for few values of $\omega_T/\omega_L$. From the experimental point of view, for Rb$^{87}$ Fig.~\ref{resonance} 
displays the life-time $2\pi/\Gamma$  (solid black line) and the gradient $G$ (red dashed line)
as functions of the bias field along the $z$ axis, for the four different values of 
$\omega_T/\omega_L$ depicted in Table~\ref{exptab}. For instance, for a value of 
$\omega_T/\omega_L = 0.2$ (top-left plot), for a bias field value of $5\, \mu T$ , the 
life-time is $\approx70$ms, whereas $G$ is $\approx25 T/m$. For these values, the trap 
frequency is $\approx14.5$kHz. The energy splitting between the $\pm 1$ states is $2.2$kHz.
The splitting of the energy levels can be measured using RF-spectroscopy on the magnetic 
trap for a cold thermal cloud \cite{Martin1988}.

\begin{table}[!ht]
\begin{ruledtabular}
\begin{tabular}{|c|c|c|c|c|}
 $\omega_T/\omega_L$  & 0.2 & 0.25 & 0.31 & 0.4 \\
 \hline
 $\Gamma_{\text{num}}\times10^{3}$ & $1.0$ & $4.8$ & $16.0$ & $41.8$\\
 \hline
$E^-_{\text{num}}-E^+_{\text{num}}$ & 0.15 & 0.18 &  0.21 & 0.25\\
\end{tabular}
\end{ruledtabular}

\caption{Decay rate of the $m=0$ ground state and the energy splitting between the
first two $m=\pm1$ states for four values of $\omega_T/\omega_L$.}
\label{exptab}
\end{table}

\section{Conclusions}
We have shown that a tight magnetic trap allows for investigating the break-down  of the Born-Oppenheimer 
condition. For instance,  the adiabatic corrections  reduce trap frequencies and the Born-Huang 
potential counteracts the adiabatic potential. The coupling between trapping and anti-trapping states results in losses which, nevertheless remain experimentally tolerable. In addition, we have shown how molecular Aharonov-Bohm gauge potentials responsible for topological Berry phase effects are emerging. These effects are two-dimensional analogues of the Weyl cones that have been simulated in cold atoms \cite{Suc06}.
Within experimentally accessible parameter ranges, i.e reasonable $B_z$,  one
could measure Majorana losses and the splitting between $m=\pm 1$, due to Berry connection 
and Born-Huang terms characterizing thereby their impact. 
As a future work, one could study the dynamics of a wave packet, for instance shifting  the 
trap center. Finally, it would be interesting to study the impact of atom-atom interactions. 
\section*{Acknowledgments}
The Centre for Quantum Technologies is a Research Centre of  Excellence funded by the 
Ministry of Education and National Research Foundation of Singapore. E. S. acknowledges financial support from the Swedish
Research Council (VR) through Grant No. D0413201. The project leading to this publication has received
funding from Excellence Initiative of Aix-Marseille Uni-
versity - A*MIDEX, a French “Investissements d’Avenir”
program through the IPhU (AMX-19-IET-008) and
AMUtech (AMX-19-IET-01X) institutes.


\end{document}